\documentclass{article}

\usepackage{arxiv}

\usepackage[utf8]{inputenc} 
\usepackage[T1]{fontenc}    
\usepackage{hyperref}       
\usepackage{url}            
\usepackage{booktabs}       
\usepackage{amsfonts}       
\usepackage{nicefrac}       
\usepackage{microtype}      
\usepackage{multirow} 
\usepackage{graphicx}
\usepackage{amsmath}
\usepackage{float}
\usepackage[numbers]{natbib}
\usepackage{doi}
\usepackage{authblk}
\usepackage[symbol]{footmisc}

\title{Application of the Digital Annealer Unit in Optimizing Chemical Reaction Conditions for Enhanced Production Yields}

\author[1,2]{Shih-Cheng Li$^{*}$}
\author[3,4]{Pei-Hwa Wang$^{*}$}
\author[3]{Jheng-Wei Su}
\author[1]{Wei-Yin Chiang}
\author[1]{Shih-Hsien Huang}
\author[1]{Yen-Chu Lin}
\author[5]{Chia-Ho Ou}
\author[3]{Chih-Yu Chen$^\dagger$}

\affil[1]{Insilico Medicine Taiwan Ltd., Taipei City 110208, Taiwan}
\affil[2]{Department of Chemical Engineering, Massachusetts Institute of Technology, Cambridge, MA 02139, United States}
\affil[3]{Quantum Information Center, Chung Yuan Christian University, Taoyuan City 320314, Taiwan}
\affil[4]{Master Program in Intelligent Computing and Big Data, Chung Yuan Christian University\\
Taoyuan City 320314, Taiwan}
\affil[5]{Department of Computer Science and Information Engineering, National Pingtung University\\
Pingtung City 900392, Taiwan}
\affil[*]{These authors contributed equally to this work.}
\affil[$\dagger$]{Correspondence: \texttt{chihyuqc@cycu.edu.tw}}

\begin{document}
\maketitle

\begin{abstract}
Finding appropriate reaction conditions that yield high product rates in chemical synthesis is crucial for the chemical and pharmaceutical industries. However, due to the vast chemical space, conducting experiments for each possible reaction condition is impractical. Consequently, models such as QSAR (Quantitative Structure-Activity Relationship) or ML (Machine Learning) have been developed to predict the outcomes of reactions and illustrate how reaction conditions affect product yield. Despite these advancements, inferring all possible combinations remains computationally prohibitive when using a conventional CPU. In this work, we explore using a Digital Annealing Unit (DAU) to tackle these large-scale optimization problems more efficiently by solving Quadratic Unconstrained Binary Optimization (QUBO). Two types of QUBO models are constructed in this work: one using quantum annealing and the other using ML. Both models are built and tested on four high-throughput experimentation (HTE) datasets and selected Reaxys datasets. Our results suggest that the performance of models is comparable to classical ML methods (i.e., Random Forest and Multilayer Perceptron (MLP)), while the inference time of our models requires only seconds with a DAU. Additionally, in campaigns involving active learning and autonomous design of reaction conditions to achieve higher reaction yield, our model demonstrates significant improvements by adding new data, showing promise of adopting our method in the iterative nature of such problem settings. Our method can also accelerate the screening of billions of reaction conditions, achieving speeds millions of times faster than traditional computing units in identifying superior conditions. Therefore, leveraging the DAU with our developed QUBO models has the potential to be a valuable tool for innovative chemical synthesis. 
\end{abstract}


\section{Introduction}
\label{sec:intro}

Organic synthetic chemistry, a field of study that focuses on creating organic compounds, is widely applied in the chemical industry and pharmaceuticals. Traditionally, synthesizing target compounds required expert chemists to design synthetic routes, along with reasonable experimental settings, such as the choice of solvents, enzymes, additives, concentration, and temperatures. However, given the complexity of chemical space, it is often difficult to identify some of the optimal reaction scenarios which lead to higher product yield. Given the nature of the complicated interaction of different factors, how to appropriately choose the conditions for a synthesis remains a significant challenge, as even minor changes in these factors can affect the final product yield. For example, some metal-catalysed reactions such as Suzuki and Buchwald-Hartwig cross-coupling reactions that are broadly used in numerous areas of applied research, are significantly influenced by the selection of catalysts and ligands. Since the synthesis of a molecule often involves multiple chemical reactions, any single reaction with a low yield within the process can contribute to overall target yield loss. Thus, developing a prediction model capable of accurately predicting reaction yields under diverse reaction conditions would be an invaluable tool, enabling chemists to select optimal reaction conditions and accelerate the invention of novel molecules.

The rapid development of machine learning brings broader applications, including chemical property prediction \cite{wang2019pgprules, heid2023chemprop, li2024when}, molecular docking \citet{crampon2022machine, stark2022equibind}, and retrosynthesis \cite{yu2023machine, coley2019robotic, koscher2023autonomous}.
A common issue in this field is data scarcity, particularly in \textit{de novo} molecular design, where the dataset often comprises only a few hundred data points. Although it has recently been shown that adding some target-related descriptors can potentially improve model performance, \cite{li2024when} intelligently selecting the subsequent experiment to be conducted is crucial. Even a modest improvement in model accuracy with each iteration can cumulatively result in significant advancements. Reker et al. \cite{reker2020adaptive} employed an active learning approach within their framework to optimize reaction conditions during synthetic procedures. Initially, they selected 5-10 data points with randomly chosen reaction conditions to train a random forest model. This model then predicted the most promising reaction condition for subsequent experimentation. Each predicted condition was evaluated, and the findings were used to inform the next cycle of model training and prediction. This active learning framework yielded superior outcomes compared to those determined by the researchers through conventional methods. Similar active machine learning methods have been extensively used for reaction yield predictions in recent yields. \cite{eyke2020iterative, shim2022predicting} These methods can significantly streamline the experimental process by identifying the most informative experiments to conduct from a potentially vast set of possibilities. In addition to active learning, reinforcement learning is also applied in the search for the most general applicable conditions. Wang et al.\cite{wang2024identifying} utilize the multi-armed bandit algorithm based on the few known reaction conditions and yields to iteratively search for the generally applicable reaction condition that brings the highest average yield on different substrates. The model iteratively updates the chance to explore the unknown condition or select the seen conditions to optimize the reaction yield. The method is examined by applying it to the existing data, and the general condition can be found with 2\% of the whole training data. Despite recent progress, screening millions or billions of combinations remains computationally demanding on traditional computers. Thus, optimizing reaction conditions remains a problem that has not been addressed.

The task of identifying optimal reaction conditions from a finite set of options is inherently a combinatorial problem. In quantum computing, several quantum-inspired algorithms are particularly adept at navigating these complex search spaces to pinpoint the best solution. Notable examples include the Quantum Approximate Optimization Algorithm (QAOA), Variational Quantum Eigensolver (VQE), and digital annealer. Among these, digital annealer is most suitable for combinatorial optimization problems, such as delivery planning involving various constraints. The main advantage of digital annealers over classical optimization methods is their potential to explore the solution space more efficiently. They utilize principles of quantum mechanics, such as quantum tunneling and entanglement, to transcend the limitations of classical algorithms. This enables them to rapidly investigate multiple possibilities simultaneously and escape local minima more effectively, while classical algorithms often get trapped in local minima. Consequently, for problems that are particularly complex and have rugged energy landscapes, digital annealers can offer significant computational benefits.

Inspired by quantum annealing techniques, Digital Annealing Unit (DAU) \cite{matsubara2020digital} was proposed to perform similar tasks of searching for optimal solutions on a global scale with enhanced stability. The target problem is first translated into the form of Quadratic Unconstrained Binary Optimization (QUBO) form; the DAU executes the digital annealing process to locate the global minimum state and determine the optimal solution. The DAU has been employed for stable structure calculations of middle-molecule drug candidates.\cite{matsubara2020digital} This molecule, comprised of 48 amino acids, is coarse-grained and modeled using the Ising model. The DAU is capable of identifying the molecule structure with minimal energy within one minute. Furthermore, DAU also helps drug discovery on lead optimization. According to Snelling et al., \cite{snelling2020quantum} it has been used to search for the optimal combination of fragments in the fragment-based chemical library to form the molecule whose pharmacophore can be mapped to the target binding sites with desired drug properties.

In this study, we integrate ML and quantum annealing to develop two QUBO models for optimizing chemical reaction conditions. The first model, an ML-based approach, uses a second-order polynomial to fit training data, while the second, a DAU-based model, assesses each reaction condition’s impact in a binary decision-making framework with binary-encoded reaction yields. Tested against held-out datasets from four high-throughput experimentation (HTE) setups, both models demonstrated their ability to accurately predict reaction yields. Additionally, the generalizability of the models and their performance with an active learning strategy were also examined. These findings support a proposed hybrid workflow that uses ML to train a model and adopts a DAU during inference. This approach can both enhance prediction accuracy and streamline the optimization process in practical chemical applications, demonstrating potential in tackling the complex challenges of optimizing reaction conditions.

\section{Methods}\label{sec:Methods}

\subsection{Data Preparation}
In this study, we utilize a range of datasets, varying in size, encompassing both high-throughput experimentation (HTE) and data extracted from the Reaxys database. The HTE datasets focus on several reactions, including Buchwald–Hartwig C–N cross-coupling \cite{ahneman2018predicting}, C–H arylation \cite{wang2024identifying}, amide coupling \cite{wang2024identifying}, and deoxyfluorination \cite{nielsen2018deoxyfluorination}. Specifically, the Buchwald–Hartwig C–N cross-coupling HTE dataset included 300 unique aryl halide and isoxazole additive combinations, along with 4 palladium pre-catalyst ligands and 3 organic bases, aggregating to 3600 experiments. In the case of the palladium-catalyzed imidazole C–H arylation reaction, the dataset involves 64 substrate pairings combined with 24 ligands, yielding 1536 data points. For the amide coupling reactions, the dataset encompasses a combination of 10 anilines, 4 bases, 3 solvents, and 8 activators, resulting in a total of 960 experiments. The dataset for deoxyfluorination reactions includes a mixture of 4 bases and 5 sulfonyl fluorides, leading to a total of 740 experiments.      
On the other hand, three reaction families are extracted from the Reaxys database by searching for their respective reaction names. Those reactions include Negishi reactions, Buchwald-Hartwig C-N cross-coupling reactions, and Suzuki reactions. Each dataset undergoes a series of preprocessing steps:
\begin{itemize}
  \item[i.] Remove reactions without reported yield, solvent, reaction SMILES, or temperature.
  \item[ii.] Remove reactions without recorded structures (half reactions).
  \item[iii.] Remove reactants or products that cannot be parsed by RDKit \cite{rdkit}.
  \item[iv.] Remove reaction conditions involving more than one product, more than two solvents, or four reagents, those without reagents, or with multiple yield records.
  \item[v.] Retain only the maximum temperature from multiple temperature records, as reactions typically occur at high temperatures. Reactions with maximum temperature lower than 20 \textdegree{}C or higher than 150 \textdegree{}C are excluded. The temperature data is further processed by:
  \begin{itemize}
  \item classifying into evenly distributed bins 
  \item encoding in binary format
  \end{itemize}
  \item[vi.] Standardize reagent and solvent labels using OPSIN \cite{lowe2011chemical}, PubChem \cite{wang2009pubchem}, and ChemSpider \cite{pence2010chemspider} to obtain SMILES. Merge labels with identical SMILES; keep labels without corresponding SMILES unchanged.
  \item[vii.] Remove duplicate reaction condition records.
\end{itemize}
In this study, we do not differentiate between catalysts and reagents as recorded in the Reaxys database; instead, these two categories are combined into one, which is referred to as "reagents." Another notable ambiguity arises when certain labels are found to reference identical chemical compounds with distinct names in different data records. To address this issue, OPSIN, PubChem, and ChemSpider are used to identify the canonical SMILES of a given compound. The most frequently used chemical name is thus used as the label for this compound. For the representation of temperature data, two methods are employed: (a) classifying into evenly distributed bins and (b) encoding in binary format. In the first method, quantile-based discretization is utilized to partition variables into uniform-sized intervals. The number of quantiles used in each dataset is 6, 9, and 9, respectively. It should be noted that this approach results in temperature ranges within bins, potentially introducing challenges in determining precise temperatures. Conversely, the second method involves encoding values in binary form, offering the advantage of precise temperature representation.
The original dataset comprises 14,025 Negishi reactions, 49,182 C-N cross-coupling reactions, and 293,083 Suzuki reactions. Following preprocessing, each dataset retains 5,188, 15,474, and 94,920 reactions, respectively. 

\subsection{Quadratic Unconstrained Binary Optimization (QUBO)}
In this work, DAU is used to optimize reaction conditions to maximize production yields since it is difficult to solve with existing general-purpose computers. In order to use DAU, a problem needs to be formulated as a QUBO form. The mathematical form is expressed as:
\begin{equation}
    y = \mathbf{x}^T Q \mathbf{x} = \sum_{i=1}^{n} \sum_{j=1}^{n} Q_{ij} x_i x_j, 
    \label{equ1}
\end{equation}
where $\mathbf{x}$ denotes a vector with its components $x_i$ being either 0 or 1, $Q$ is a $n$-dimensional square matrix. Each element $Q_{ij}$ of this matrix signifies a weight corresponding to each pair of indices $i$, $j$ within the vector $\mathbf{x}$. In the context of optimizing reaction yield of chemical reaction, the vector $\mathbf{x}$ encapsulates reaction information such as reactants, solvents, and reagents. The matrix $Q$ functions as a mathematical model, designed to evaluate the yield of the chemical reaction under various conditions represented by the vector $\mathbf{x}$. This model is particularly adept at mapping out how different combinations of reaction conditions influence the reaction yield.

To optimize reaction yields, the objective function is defined as minimizing $-\mathbf{x}^T Q \mathbf{x}$, effectively identifying the conditions under which the reaction yield is optimized. The binary nature of $\mathbf{x}$ implies that each element of the vector represents the presence (1) or absence (0) of a particular component in the reaction, thereby allowing for a discrete optimization approach.

Equation \ref{equ1} represents a QUBO problem in its basic form, without explicit constraints. Nonetheless, practical applications often necessitate adherence to certain constraints as the optimizer seeks optimal solutions. For instance, in some reactions, no more than one base is permitted. Most of such constraints can be re-formulated into QUBO form by embedding quadratic penalties within the objective function. \cite{glover2022quantum} This method assigns a zero penalty for solutions that meet the constraints (feasible solutions) and imposes a positive penalty on those that violate them (infeasible solutions). As a result, the optimizer is directed to minimize an enhanced objective function. Ideally, this augmented function should align with the original objective function when the imposed penalties become negligible or are effectively nullified.

\subsection{QUBO Model for Reaction Yield Prediction}
Defining a QUBO form for optimizing reaction yields is complex due to the inherent complexity of a chemical reaction. \cite{marin2019kinetics} It is rooted in various factors, including the nature of reactants and products, reaction pathways, and mechanisms. Key elements such as activation energy, the role of catalysts, and environmental conditions like temperature and pressure significantly influence the reaction's course and speed. Additionally, concepts like chemical equilibrium, reaction kinetics, and thermodynamics play vital roles in determining the outcomes and efficiencies of reactions. These factors, combined with considerations of potential by-products, underscore the intricate and multifaceted nature of chemical reactions.

\begin{figure}[h!]
    \centering
    \includegraphics[width=0.8\linewidth]{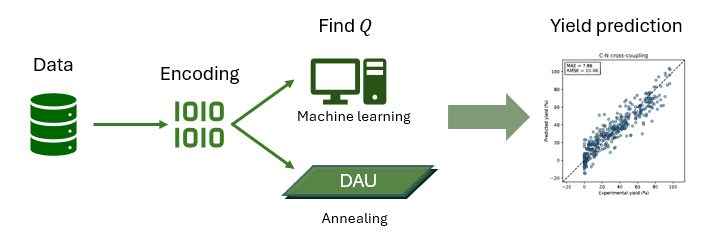}
    \caption{Workflow of this work. The reaction data from HTE/Reaxys are encoded by one-hot encoding/fingerfrints. The $Q$ matrix is then obtained through an ML-based/DAU-based model. With the $Q$ matrix, one can use it to predict reaction yields from a given reaction information.}
    \label{fig:workflow}
\end{figure}

Due to the complexity, designing a general QUBO formulation for reaction yield optimization is generally impossible. Therefore, we use ML to determine the $Q_{ij}$ coefficients in Equation \ref{equ1}. Since $Q_{ij}$ is a learned parameter, the model performance is shaped by three main elements: the method of reaction encoding, the number of data points, and the chemical space encountered during the training phase. In this work, as illustrated in Figure \ref{fig:workflow}, we compare the performances of the ML-based method with the DAU-based counterpart, which learns the reaction information directly through the mechanism of a digital annealer.

\subsubsection{ML-based Model}\label{sec:ml_qubo}
For reaction encoding, two methods are utilized: one-hot encoding and a hybrid approach combining one-hot encoding with fingerprint representations. The encoding methods for each dataset and its possible application are summarized in Table \ref{table:encoding_summary}.%

\begin{table}[]
\caption{Encoding method and its possible application in this work. }
\begin{center}
\begingroup
\renewcommand{\arraystretch}{1.3} 
\resizebox{\textwidth}{!}{

\begin{tabular}{|c|cccl}
\hline
\multicolumn{1}{|c|}{Data}          & \multicolumn{2}{c|}{Encoding}              & \multicolumn{2}{c|}{Possible Application} \\ 
\hline\hline
\multicolumn{1}{|c|}{}  & \multicolumn{1}{c|}{Reactant} & \multicolumn{1}{c|}{Reaction condition} & \multicolumn{2}{c|}{} \\ 
\hline
\multicolumn{1}{|c|}{\multirow{2}{*}{HTE dataset}} & \multicolumn{1}{c|}{Bin-based one-hot encoding}          & \multicolumn{1}{c|}{\multirow{3}{*}{Bin-based one-hot encoding}} & \multicolumn{2}{c|}{Yield/condition prediction for the known reactant}     \\ 
\cline{2-2} \cline{4-5} 
\multicolumn{1}{|c|}{}   & \multicolumn{1}{c|}{Substructure-based one-hot encoding} & \multicolumn{1}{c|}{}   & \multicolumn{2}{c|}{Yield/condition prediction for the known substructure} \\ 
\cline{1-2} \cline{4-5} 
\multicolumn{1}{|c|}{Reaxys dataset}  & \multicolumn{1}{c|}{Fingerprints(MACCS/Avalon/ECFP4)}   & \multicolumn{1}{c|}{}    & \multicolumn{2}{c|}{Yield/condition prediction for the known reactant in a given reaction type}  \\ 
\hline
\end{tabular}
}
\endgroup
\end{center}
\label{table:encoding_summary}
\end{table}

For the one-hot encoding approach, reaction conditions such as solvents, bases, or additives are encoded as separate binary indicators. This method is effective because there are usually only dozens or hundreds of species are considered, and their chemical structures are significantly different. One-hot representation is also applicable in encoding reactants, particularly in a smaller chemical space. This can be done in two ways: encoding each species as a separate bin or based on its substructures. The bin-based approach faces limitations when introducing a new reactant. In such cases, the QUBO model needs to be reformulated by increasing the size of $x$ and $Q$ in Equation \ref{equ1} by 1. Alternatively, if all relevant reactants can be broken down by their substructures (e.g., molecular backbones, functional groups), the model can adapt to new species through the presence or absence of specific substructures. However, predictions for these novel species are considered extrapolation tasks, and the accuracy of these predictions must be rigorously evaluated before applying the model.
As for the hybrid approach, the reactants or products are encoded by expert-guided fingerprints, including 166-bits MACCS \cite{durant2002reoptimization}, 1024-bits Avalon \cite{gedeck2006qsar}, and 2048-bits ECFP4 \cite{rogers2010extended} fingerprints, while the reaction conditions are still encoded by one-hot representation. This hybrid approach is recommended for use when the chemical space being considered is relatively large, which results in a fingerprint size smaller than that obtained with one-hot encoding. In this study, this method is particularly suitable for application to Reaxys datasets, where the number of unique reactants is thousands. Especially, the majority of these reactions are associated with only a single reaction condition, which makes the use of one-hot encoding impractical, as it fails to provide any new knowledge for new species.

For predicting reaction properties, alterations in local structure are highly informative. They provide insights into the mechanisms of chemical reactions and the reorganization of atoms and bonds. Various approaches can be adopted to leverage this information as model input. One option is to use a straightforward concatenation of the representations of reactants and products (reac\_prod), or to focus solely on the difference between reactants and products (diff\_only). Additionally, this difference can be concatenated to the representation of either the reactants (reac\_diff) or the products (prod\_diff). Since individual models can be developed for each reaction family, it's possible to predict the product based on the reactants, and vice versa. Consequently, it's also feasible to use only the information from either the reactants (reac\_only) or the products (prod\_only). In this study, all used fingerprints are binary representations. However, the difference between two such fingerprints yields three values: -1, 0, and 1, corresponding to the disappearance, unaltered state, and addition of a specific substructure within a particular reaction, respectively. Concerns regarding the compatibility of the QUBO model with such inputs may arise. While it is correct that using ML to find the $Q$ matrix does not adhere to the QUBO's inherent reliance on binary representations, there is no restriction on whether the input should be binary when using CPU or GPU to train this model. The situation differs during the inference phase on DAU. Thus, we integrate the reaction information into the re-formulated $Q$ matrix, which adheres to the binary fusion of inputs. This is done by first obtaining the polynomial expression of $x^T Q x$, followed by converting it into quadratic models of the QUBO form.

\subsubsection{DAU-based Model}

To predict the yield of a reaction in QUBO form, we assign 10 binary variables to every reaction condition and the pairwise combinations of conditions with coefficients ranging from $2^0$ to $2^9$. Considering unfavorable conditions may decrease the product yield, the range is adjusted to [-512, 511]. For example, the contribution of the first reaction condition is:\\
\begin{equation}
    2^0x_0+2^1x_1+2^2x_2+2^3x_3+2^4x_4+2^5x_5+2^6x_6+2^7x_7+2^8x_8+2^9x_9-512 
    \label{eq:binary_encoding}
\end{equation}

Equation \ref{eq:binary_encoding} serves as the contribution of either a condition or a certain pairwise combination of conditions in yield prediction. The prediction of a reaction is the sum of the contribution of every condition and pairwise combination of conditions. For example, if there are two reaction conditions in Reaction 1, the yield prediction can be calculated as below.\\
\begin{itemize}
  \item Contribution of substrate 1:\\
  \begin{equation}
    s_1=2^0x_0+2^1x_1+2^2x_2+2^3x_3+2^4x_4+2^5x_5+2^6x_6+2^7x_7+2^8x_8+2^9x_9-512
    \label{eq:binary_encoding_s}
  \end{equation}
  \item Contribution of additive 1:\\
  \begin{equation}
    a_1=2^0x_{10}+2^1x_{11}+2^2x_{12}+2^3x_{13}+2^4x_{14}+2^5x_{15}+2^6x_{16}+2^7x_{17}+2^8x_{18}+2^9x_{19}-512
    \label{eq:binary_encoding_a}
  \end{equation}
  \item Contribution of pairwise combination of substrate 1 and additive 1: \\
  \begin{equation}
   s_1a_1=2^0x_{20}+2^1x_{21}+2^2x_{22}+2^3x_{23}+2^4x_{24}+2^5x_{25}+2^6x_{26}+2^7x_{27}+2^8x_{28}+2^9x_{29}-512
   \label{eq:binary_encoding_sa}
  \end{equation}
  \item Prediction yield of the reaction 1:\\

  \begin{equation}
  P_{1}=s_1+a_1+s_1a_1
  \label{eq:binary_encoding_p}
  \end{equation}
\end{itemize}

The objective function is the sum of squared error of training set and validation set combined together.\\
\begin{align}
    Objective &=\sum_{train}(P_i-Y_i)^2 \label{eq:binary_encoding_obj}
\end{align}
Where $s$, $a$, $P$, and $Y$ are substrate, additive, production yield, and the ground truth value (i.e., experimental yield), respectively.

The result of DAU is the solution of vector $x$ that can minimize the objective function, which is similar to minimizing the square error in training the predictive model. We choose the top-1 solution reported by DAU to build models in this study. The coefficients in the matrix $Q$ in the QUBO problem $x^TQx$ are the contribution of choosing a reaction condition (diagonal elements) and pairwise combinations of conditions (other elements), which can be obtained by running DAU to find the least squares fitting solution.

\subsection{Model Training}
\begin{figure}
     \centering
     \includegraphics[width=0.9\linewidth]{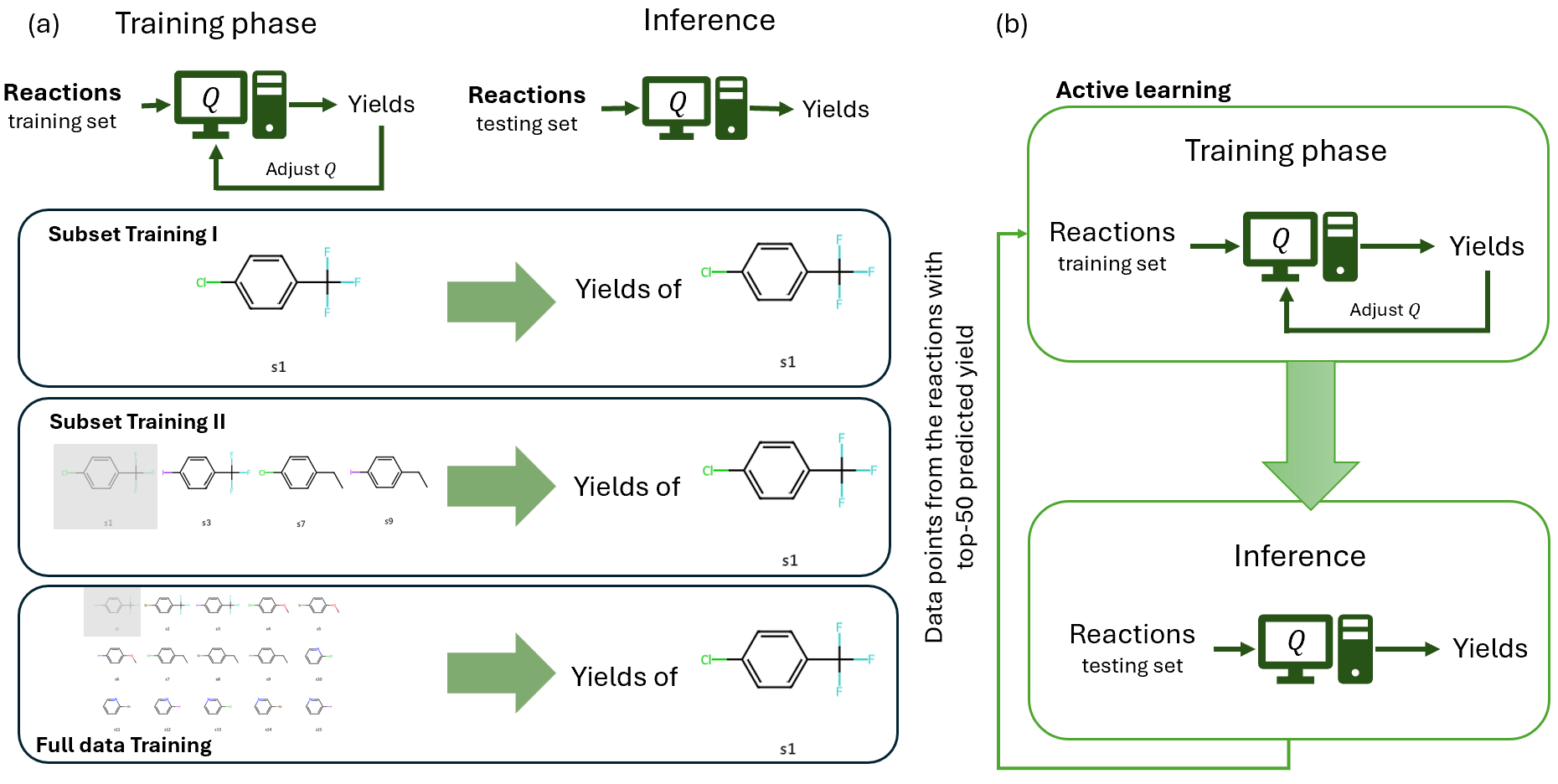}

        \caption{Schematic illustration of (a) subset/whole data training and (b) active training. In subset training I, the subset data is used for training and predicting the yield of the corresponding subset, while in subset training II, the model's extrapolation ability is verified by training on three subsets and testing on the remaining one. For active training in (b), 100 data points are randomly sampled from each HTE dataset initially. 50 data points are either randomly or strategically added to the training data iteratively.}
        \label{fig:training_procedure}
\end{figure}

In this work, a couple of different models are trained, including (1) models trained on subsets, (2) models trained on full datasets, and (3) models with active learning, as illustrated in Figure \ref{fig:training_procedure}.

In the first task, the C–N cross-coupling HTE dataset is divided into subsets based on the substrates used. This dataset includes a combination of 15 substrates, 4 ligands, 3 bases, and 20 additives, encompassing a total of 3,600 data points. Subsets are created with s1, s3, s7, and s9 as their respective substrates. Each subset comprises 240 data points. The aim of training on these subsets is to determine whether ML-based QUBO formulation can accurately link the input with the target, compared to some commonly seen ML methods such as feed-forward neural network (FFN), Support Vector Machine (SVM), and Random Forest (RF). The 15 substrates in this dataset are one-hot encoded into 8 bits based on their substructures (Figure \ref{fig:cn-processed_substrate}). Since the substructure in any of s1, s3, s7, and s9 also appear in the other 3 substrates, it is intriguing to assess the model's extrapolation ability by training on three subsets and testing on the remaining one. Each dataset is split in an 80:10:10 ratio for the training, validation, and testing sets.

For the models trained on full datasets, each dataset (i.e., four HTE datasets and three Reaxys datasets) is randomly partitioned into 80\% training data, 10\% validation data, and 10\% test data for model training. For HTE datasets, except for the C-N cross-coupling dataset, the encoding of reactants is based on the bin-based method. For Reaxys datasets, three kinds of fingerprints (i.e., MACCS, Avalon, and ECFP4) are used.

In the active learning experiments, 100 data points are randomly sampled from each HTE dataset in five folds, and each training set is expanded iteratively, increasing its size by 50 each time. Two methods are proposed for including new data: (1) random selection of 50 data points, and (2) strategic selection of the top 50 data points with higher prediction yields. To study the performance of the proposed algorithm, the loop is stopped when the number of molecules in the training set reaches 550, equating to the addition of nine new batches of data.

Each ML-based QUBO model is trained for 200 epochs with an ensemble size of 5 and a learning rate of $10^{-3}$. The prediction is obtained by averaging the outputs of the submodels in the ensemble. The targets are normalized to have zero mean and unit standard deviation. For the FFN model, the number of layers is 2 and the number of hidden neurons is 300. The number of trees used in the RF model is 500. We did not perform hyperparameter optimization here, as the aim of training these models was to compare whether the QUBO model can achieve similar performance to traditional ML methods. 

In DAU-based model construction, Fujitsu DA3c was used and the default setting was applied. The running time of DA is set to ten seconds. DA reported the best solutions found in the ten-second annealing process. In average, the best solution was found in seven to nine seconds in the active learning experiments and the discovery time decreased gradually with iterations. The best solution outputted from DAU was used in the study.

\section{Results and Discussion}

\subsection{Evaluation of ML- and DAU-based Models on C-N cross-coupling Yield Prediction}

\begin{figure}[h]
  \centering
  \includegraphics[width=0.9\linewidth]{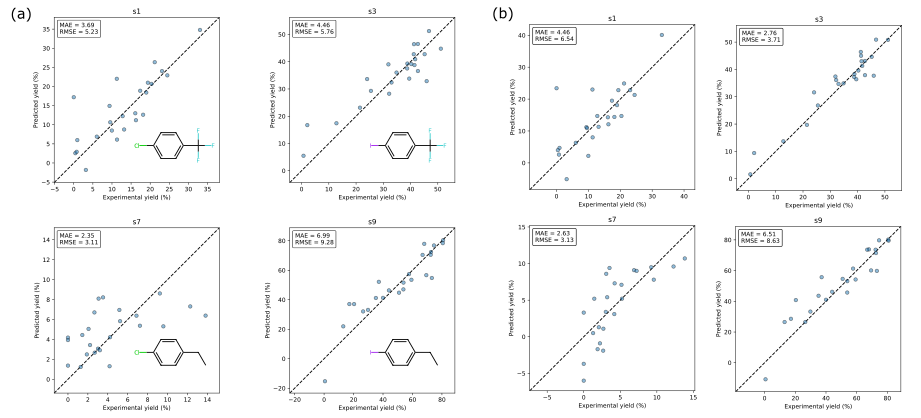}

  \caption{The parity plots of experimental yields and the corresponding predicted yields from the (a) ML-based model and (b) DAU-based model on the test sets. These parity plots of the ML-based model were separately trained on s1, s3, s7, and s9 subsets of the C-N cross-coupling dataset, with one-hot encoding applied to the agents. The strong dependencies between the predicted and experimental yields show the efficacy of QUBO model in predicting reaction yields. (HTE datasets)}
  \label{fig:subsets_parity_plot} 
\end{figure}

To examine where the QUBO model can be used as an estimator to predict the reaction yield of a reaction under various reaction conditions, we first choose some subsets (s1, s3, s7, and s9) from the C-N cross-coupling dataset based on the participating substructures. In Figure \ref{fig:subsets_parity_plot}, a strong correlation is observed between the actual and predicted reaction yields for most cases, suggesting the efficacy of the QUBO model in predicting reaction yields. The considerable deviations in outliers from the reference values are attributed to their locations in regions that are underrepresented in the corresponding subset.

The same data splits are used to train the FFN, SVM, and RF baseline models for comparison purposes. Table \ref{table:baselines} shows that our model performance is comparable to both the FFN and RF, and outperforms the SVM model. It is important to know that using either FFN or RF might be computationally expensive for inferring datasets containing millions of data points. As mentioned in the Methods Section, QUBO is designed specifically for large-scale combinatorial optimization problems. Although it is possible neural network methods are more suitable for more complex problems, the ease of interpretation, explanation, and speed of inference with digital annealers are notable advantages of this model.

\begin{table}[h]
\caption{Comparative analysis of QUBO and baseline models on C-N cross-coupling dataset subsets s1, s3, s7, and s9 (in HTE datasets). Evaluation metrics include MAE and RMSE. The comparisons show the performance of the QUBO models are compatible with FFN and RF and outperforms the SVM model. (HTE datasets)}

\begin{center}
\begingroup
\renewcommand{\arraystretch}{1.3} 
\resizebox{0.9\textwidth}{!}{%
\begin{tabular}{lcccccccc} 
\toprule
& \multicolumn{2}{c}{s1} & \multicolumn{2}{c}{s3} & \multicolumn{2}{c}{s7} & \multicolumn{2}{c}{s9} \\
\cmidrule(lr){2-3} \cmidrule(lr){4-5} \cmidrule(lr){6-7} \cmidrule(lr){8-9}
Model & MAE & RMSE & MAE & RMSE & MAE & RMSE & MAE & RMSE \\
\midrule

ML-based QUBO & 3.69 & 5.23 & 4.46 & 5.76 & 2.35 & 3.11 & 6.99 & 9.28 \\
DAU-based QUBO & 4.46 & 6.54 & 2.76 & 3.71 & 2.63 & 3.13 & 6.51 & 8.63 \\
FFN & 3.43 & 4.92 & 3.54 & 4.85 & 2.68 & 3.39 & 5.14 & 6.64 \\
SVM & 6.46 & 7.53 & 7.97 & 11.98 & 2.38 & 3.24 & 17.06 & 22.87 \\
RF & 4.53 & 5.70 & 3.75 & 5.52 & 2.91 & 4.04 & 7.06 & 9.59 \\
\bottomrule
\end{tabular}%
}
\endgroup
\end{center}
\label{table:baselines}
\end{table}

To assess the extrapolation ability of our model when using substructure-based one-hot encoding, our model is trained on three subsets selected from s1, s3, s7, and s9, and then tested on a held-out test set to evaluate its predictive performance on unseen but correlated molecules. As illustrated in Figure \ref{fig:extrapolation_parity_plot}, the correlation of ranking for predictions and real values is high for the model tested on s3 or s9 subsets (Table \ref{table:extrapolation_result}), meaning that the model could recommend reaction conditions corresponding to high yields in real experiments. However, a significant prediction error is observed in each model. The same experiments conducted on baseline models (MLP, SVM, and RF) can be found in Table \ref{table:baselines_extrapolation} and Figures \ref{fig:extrapolation_parity_plot_MLP} to \ref{fig:extrapolation_parity_plot_random_forest}. The results also reveal a high correlation between predicted and true values for s3 and s9 subsets in MLP and RF baseline models.

For the models trained on the s1 and s7 subsets, a significant prediction error is observed, which can primarily be attributed to insufficient information on new substructure pairs and partly to uneven data distribution. When the test set contains fewer high-yield value data and exhibits a smaller average yield value compared to the training sets, such as in the cases of s1 or s7 as the test set, as shown in the Table \ref{table:statistics}, the model tends to predict higher yield values. Consequently, the prediction may not correspond accurately to the test set due to these disparities in data characteristics. Given the limited number of data used to train the model, it faces challenges in extrapolating to unseen data. An improvement in extrapolation ability is found when more diverse and substantial data are added to the model, as depicted in Figure \ref{fig:extrapolation_parity_plot_1}. As one might expect, the model performance can be significantly improved by adding data related to the new substrates, as shown in Figure \ref{fig:extrapolation_parity_plot_2}. The above experiments indicate that with a limited number of training data points, gathering additional data to train a new model is advantageous, if feasible. In cases where this is not possible, conducting pilot studies prior to using the model is suggested. These studies should include an analysis of the differences between the molecules intended for inference and those in the training data, as well as an examination of the target ranges for both the training and inferencing data. It is unrealistic to expect accurate predictions from the model regarding the range of values if these are underrepresented in the training dataset.

\begin{table}[h]
\caption{Results for the models trained on three subsets selected from s1, s3, s7, and s9 in HTE datasets, evaluated on the held-out test set of the C-N cross-coupling dataset with one-hot encoding via ML-based and DAU-based model. Top-k score and Pearson correlation coefficient are used as metrics. The top-k scores in s3, s7 and s9 show the possibility that a QUBO-based model can be used for optimal yield prediction, meaning that the model could recommend reaction conditions corresponding to high yields in real experiments.}

\begin{center}
\begingroup
\renewcommand{\arraystretch}{1.3} 
\resizebox{\textwidth}{!}{%
\begin{tabular}{lcccccccccccccccccccc} 
\toprule
& \multicolumn{5}{c}{s1} & \multicolumn{5}{c}{s3} & \multicolumn{5}{c}{s7} & \multicolumn{5}{c}{s9} \\
\cmidrule(lr){2-6} \cmidrule(lr){7-11} \cmidrule(lr){12-16} \cmidrule(lr){17-21}
 & ML & FFN & SVM & RF & DAU & ML & FFN & SVM & RF & DAU & ML & FFN & SVM & RF & DAU & ML & FFN & SVM & RF & DAU \\
\midrule
Top-5 accuracy score            & 0 & 0 & 2 & 1 & 0 & 1 & 1 & 0 & 3 & 0 & 1 & 1 & 0 & 1 & 1 & 2 & 2 & 1 & 0& 2 \\
Top-10 accuracy score           & 0 & 0 & 3 & 1 & 0 & 3 & 4 & 3 & 6 & 3 & 1 & 1 & 0 & 2 & 1 & 4 & 4 & 1 & 4 & 2 \\
Top-15 accuracy score           & 0 & 0 & 5 & 2 & 0 & 5 & 6 & 5 & 8 & 5 & 3 & 3 & 2 & 2 & 2 & 5 & 5 & 6 & 9 & 4 \\
Top-20 accuracy score           & 0 & 0 & 8 & 5 & 0 & 8 & 7 & 7 & 10 & 8 & 3 & 4 & 3 & 4 & 4 & 7 & 9 & 7 & 12& 7 \\
Pearson correlation coefficient & 0.09 &0.10& 0.5&0.56&-0.03&0.83& 0.84 & 0.69 & 0.82 & 0.83 & 0.34 & 0.33 & 0.22 & 0.38 & 0.28 & 0.72 & 0.75  & 0.58 & 0.83& 0.70 \\
\bottomrule
\end{tabular}%
}
\endgroup
\end{center}
\label{table:extrapolation_result}
\end{table}

\begin{table}[h]
\caption{Comparative analysis of QUBO and baseline models trained on three subsets selected from s1, s3, s7, and s9 in HTE datasets, evaluated on the held-out test set of the C-N cross-coupling dataset with one-hot encoding. Evaluation metrics include MAE and RMSE. The comparisons show the performance of the ML-based QUBO model is generally compatible with FFN and RF and outperforms the SVM model. (HTE datasets)}

\begin{center}
\begingroup
\renewcommand{\arraystretch}{1.3} 
\resizebox{0.9\textwidth}{!}{%
\begin{tabular}{lcccccccc} 
\toprule
& \multicolumn{2}{c}{s1} & \multicolumn{2}{c}{s3} & \multicolumn{2}{c}{s7} & \multicolumn{2}{c}{s9} \\
\cmidrule(lr){2-3} \cmidrule(lr){4-5} \cmidrule(lr){6-7} \cmidrule(lr){8-9}
Model & MAE & RMSE & MAE & RMSE & MAE & RMSE & MAE & RMSE \\
\midrule
ML-based QUBO & 13.49 & 17.65 & 15.19 & 18.53 & 28.62 & 32.20 & 31.54 & 36.12 \\
DAU-based QUBO & 88.50 & 90.07 & 37.22 & 40.32 & 70.59 & 72.71 & 49.01 & 52.42 \\
FFN & 15.81 & 19.91 & 22.69 & 25.04 & 25.92 & 29.67 & 31.30 & 35.69 \\
SVM & 15.81 & 19.91 & 12.19 & 13.32 & 27.50 & 27.95 & 34.4 & 39.37 \\
RF & 9.15 & 12.50 & 20.21 & 22.49 & 20.10 & 21.96 & 21.12 & 23.99 \\
\bottomrule
\end{tabular}%
}
\endgroup
\end{center}
\label{table:baselines_extrapolation}
\end{table}

\begin{table}[h]
\caption{The statistics of C-N cross-coupling dataset subsets s1, s3, s7, and s9 (in HTE datasets). Rows include average yield and the percentages for yield greater and less than 15\% in each subset.}

\begin{center}
\begingroup
\renewcommand{\arraystretch}{1.3} 
\begin{tabular}{lcccccccc} 
\toprule
& s1 & s3 & s7 & s9 \\
\midrule
Average Yield (\%) & 12.36 & 33.32 & 3.97 & 51.68 \\
Yield $\geq$ 15\% & 67\% & 91\% & 49\% & 94\% \\
Yield $<$15\% & 33\% & 9\% & 51\% & 6\% \\
\bottomrule
\end{tabular}
\endgroup
\end{center}
\label{table:statistics}
\end{table}

\subsection{Assessment of Model Effectiveness Across Varied Training Datasets}

The previous section confirmed that the QUBO model can serve as an alternative for predicting reaction yields. The prior tests focused solely on the C-N cross-coupling dataset. In this section, benchmarks are performed on various datasets, including the HTE and Reaxys datasets, to verify the generality of the QUBO model.

\begin{figure}[h]
  \centering
  \includegraphics[width=\linewidth]{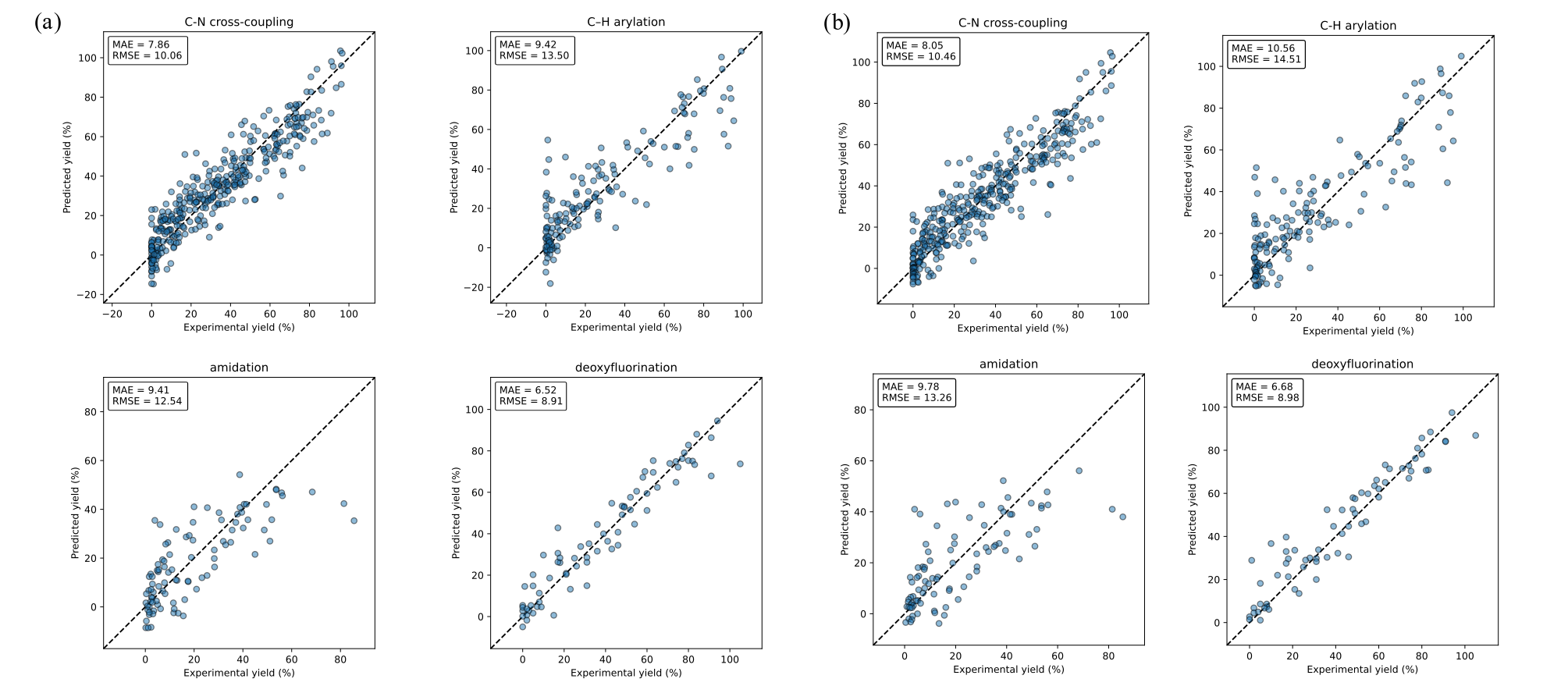}
  \caption{The results of the (a) ML-based models and (b) DAU-based models trained on the data of C-N cross-coupling, C-H arylation, amidation and deoxyfluorination in HTE dataset with one-hot encoding. The parity plots are obtained by evaluating the corresponding held-out test set in different data subsets. The correlation between predicted yield and experimenta (ground truth)) yield in each prediction shows the generality of learnability of the QUBO model.}
  \label{fig:hte_parity_plot} 
\end{figure}

Each HTE dataset provides reaction yields of every combination of reaction conditions for each reaction, whereas reactions in the Reaxys datasets usually have only one reaction condition recorded. Data uncertainty in HTE datasets is observed to be lower, as measurements are conducted under consistent experimental settings and are operated by identical research groups. Conversely, Reaxys datasets, compiled from various sources, tend to exhibit higher data uncertainty. As mentioned in the Methods section, the HTE datasets are encoded through one-hot encoding, and the Reaxys datasets are preprocessed through a hybrid encoding method. The results for models trained on HTE datasets are shown in Figure \ref{fig:hte_parity_plot} and suggest that our QUBO-based estimator can be adopted for predicting reaction yields in various types of reactions with an MAE lower than 11\% yield.

To further test the model's capability in a more complicated problem setting, we trained the ML-based QUBO model on Reaxys datasets. The results show that the testing errors are larger than those from HTE datasets, as expected, due to the increased data complexity and uncertainty. (Tables~\ref{table:negishi}, \ref{table:buchwald-hartwig}, and \ref{table:suzuki}). Even when the best combination of reaction encoding method and fingerprint is chosen, the testing MAE is still larger than 10\%. Regarding the choice of fingerprint, the MACCS fingerprint has been shown to be more suitable than the Avalon and ECFP4 fingerprints within our model. This reason might be attributed to the shorter input length of the MACCS representation. As for the methods of reaction encoding, reac\_only, prod\_only, and diff\_only have been found to be comparatively effective in our test cases, outperforming the other three approaches (reac\_prod, reac\_diff, prod\_diff). It is noteworthy that the information for the first three methods is actually included in the last three methods, albeit with twice the input length. The observed trend indicates improved model performance in current benchmarks when the input dimensionality is reduced. To substantiate this observation regarding input dimensionality, we test various bit sizes in Avalon and ECFP4 fingerprints for Negishi reactions. The findings, as shown in Table \ref{table:negishi_various_bits}, align with our initial hypothesis. Additionally, our results also suggest that temperature information does not seem important for yield prediction within our model. Theoretically, temperature influences the reaction rate, thereby affecting reaction selectivity and product yields. Our model's insensitivity to this parameter can be attributed to several factors. First, the model may be too simplistic to capture the parameter's impact. Second, selecting the maximum temperature from recorded values might be misleading, as varying temperatures could influence different processes, affecting the overall yield. Third, our encoding methods (i.e., evenly distributed bins or binary format) may hinder the model's ability to learn this information, suggesting a need for alternative encoding techniques. Fourth, the effect of temperature on the product rate is also intertwined with other parameters like reaction time and pressure. Lastly, data points in the Reaxys dataset, compiled from various sources, exhibit higher uncertainty, limiting the model's ability to utilize this information effectively.

\begin{table}[h]
\caption{Testing MAE for ML-based models trained on Negishi reactions, using various combinations of reaction encoding approaches and fingerprints, from the Reaxys database. The temperatures are either not specified, provided in evenly distributed bins, or presented in a binary format. From the results, the MACCS fingerprint has been shown to be more suitable than the Avalon and ECFP4 fingerprints within our model, while the temperature information does not seem important for yield prediction within our model.}

\begin{center}
\begingroup
\renewcommand{\arraystretch}{1.3} 
\resizebox{\textwidth}{!}{%
\begin{tabular}{lccccccccc} 
\toprule
& \multicolumn{3}{c}{Evenly distributed bins
} & \multicolumn{3}{c}{Binary encoding
} & \multicolumn{3}{c}{No contribution
} \\
\cmidrule(lr){2-4} \cmidrule(lr){5-7} \cmidrule(lr){8-10}
 & MACCS & Avalon & ECFP4 & MACCS & Avalon & ECFP4 & MACCS & Avalon & ECFP4 \\
\midrule
reac\_prod & 12.76 & 23.00 & 22.36 & 13.02 & 30.21 & 19.68 & 12.44 & 25.50 & 20.28 \\
reac\_only & 11.38 & 14.63 & 12.29 & 11.96 & 14.35 & 12.58 & 10.78 & 14.74 & 12.43 \\
prod\_only & 11.46 & 14.24 & 12.40 & 11.63 & 14.52 & 12.60 & 11.32 & 14.01 & 11.83 \\
reac\_diff & 11.13 & 26.41 & 17.29 & 11.32 & 27.50 & 18.48 & 11.72 & 27.34 & 17.20 \\
prod\_diff & 11.39 & 23.89 & 16.26 & 11.64 & 26.72 & 16.63 & 11.65 & 25.15 & 15.27 \\
diff\_only & 11.68 & 13.39 & 11.3 & 11.59 & 13.60 & 11.64 & 11.92 & 13.76 & 11.58 \\
\bottomrule
\end{tabular}%
}
\endgroup
\end{center}
\label{table:negishi}
\end{table}

\subsection{Optimizing Process Conditions Using Active Learning Techniques}

In practical applications, the identification of reaction conditions that have higher yields is a key objective for chemists. Limited data points often restrict the accuracy of predictive models. However, incremental enhancements become feasible as new data are conducted and integrated into the dataset. This raises the question, 'How can one effectively determine the nature of experiments to be conducted in subsequent iterations?' While the random incorporation of new data points may seem beneficial for model refinement, it may not be the most efficient approach. Considering the influence of data distribution on model performance, we hypothesize that enriching the dataset with higher-yield scenarios could improve model predictions regarding optimal reaction conditions. To evaluate this hypothesis, active learning experiments are conducted across various HTE datasets to explore the combination of reactants and conditions that maximize yield.

\begin{figure}[h]
  \centering
  \includegraphics[width=\linewidth]{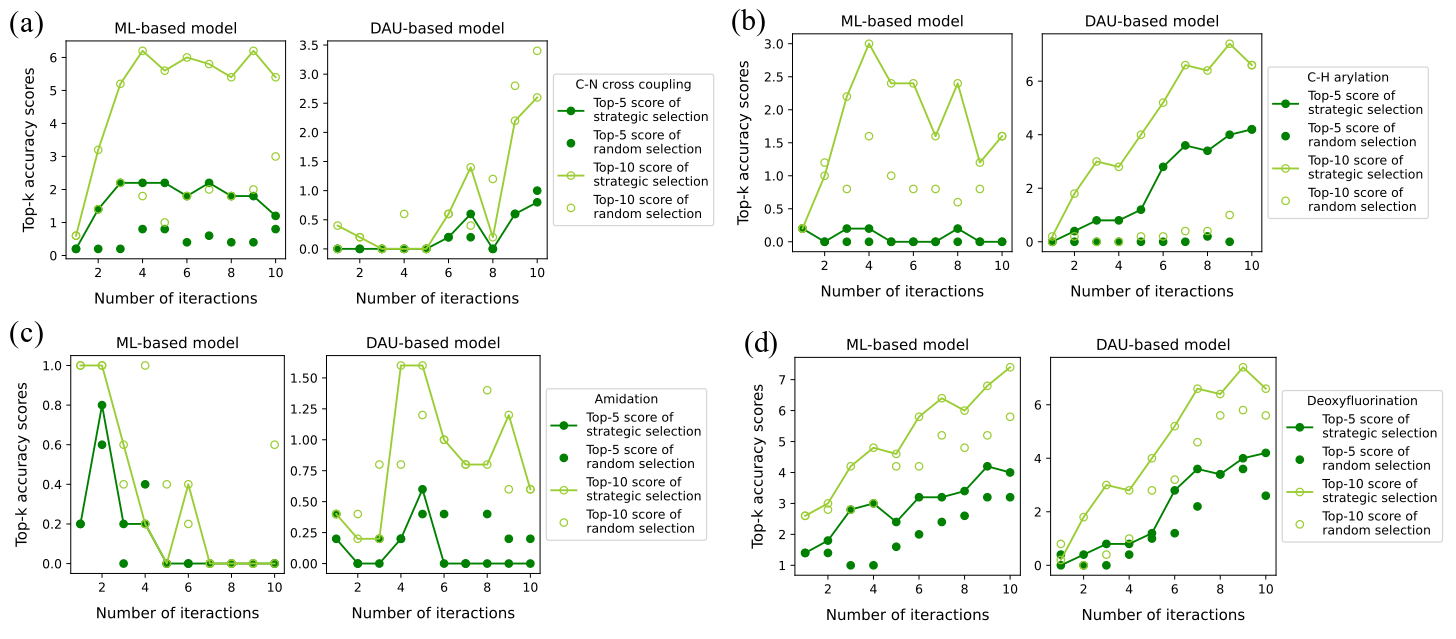}
  \caption{Results obtained from applying active learning models to diverse HTE datasets (with one-hot encoding), utilizing strategic or random methods for adding new data points in subsequent runs. Each data point denotes the mean top-k score, computed over five folds. The results show that the accuracy of active learning is saturated around the 4th or 5th iteration, and the performance of strategic selection outperforms random selection in the ML-based model. Similarly, in the DAU-based model, the performance plateaus around iterations 7 or 8, and the strategic selection also shows an advantage on optimal condition finding.}
  \label{fig:active_learning_scores} 
\end{figure}

As demonstrated in Figure \ref{fig:active_learning_scores}, strategic selection typically surpasses random selection in performance for both ML and DAU-based models. In the datasets pertaining to C-N cross-coupling and deoxyfluorination, the ML-based model is capable of identifying over half of the top-k conditions. Similarly, the DAU-based model demonstrates this capability when predicting C-H arylation and deoxyfluorination datasets. An initial phase of progressive improvement is observed in the ML-based model during the early iterations. However, this improvement tends to plateau after approximately 4 or 5 iterations. While further inclusion of data results in marginal enhancements, the magnitude of improvement is not as pronounced as observed in the initial runs. For DAU-based models, the performance plateaus after 8 iterations. Since we use ten binary variables to encode one $Q_{ij}$ coefficient in DAU-based models, the number of variables is ten times higher than in ML-based models. Thus, the model would require more data to learn the underlying chemistry, resulting in more runs to reach a plateau.

\subsection{Comparative Analysis of Computational Efficiency: Classical Computing vs. Digital Annealer Unit (DAU) in Model Inference}
\begin{figure}[h!]
    \centering
    \includegraphics[width=0.8\linewidth]{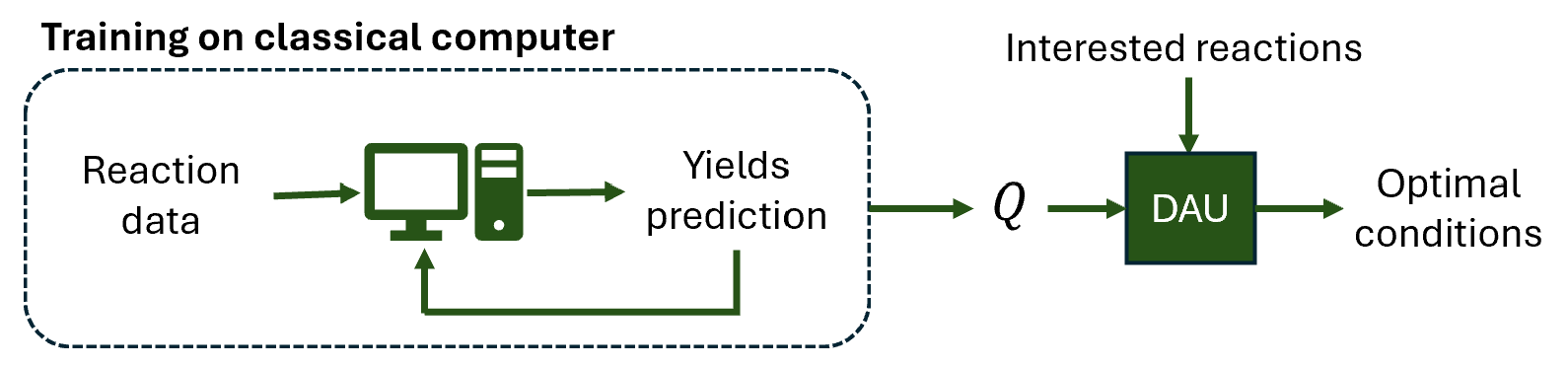}
    \caption{Workflow of the hybrid model on optimal condition search. The Q matrix is first obtained from classical machine learning, followed by the annealing process on DAU.}
    \label{fig:workflow_hybrid}
\end{figure}

One of the most pronounced advantages of an annealer is to find the optimal solution to the combinatorial problem. This advantage results from the tunneling mechanism that helps to find the minimal energy at a global scale. In some sense, this is equivalent to parallel computing that compares all possible solutions simultaneously. On the other hand, the pros and cons of ML-/DAU-based methods are realized from the results above. Based on this information, we provide a possible usage scenario, demonstrated in Figure \ref{fig:workflow_hybrid}, that utilizes the advantage of classical and quantum to achieve efficient search. The efficiency of the search procedure can be realized by the time required to obtain the optimal solution. 

\begin{figure}[h]
  \centering
  \includegraphics[width=0.5\linewidth]{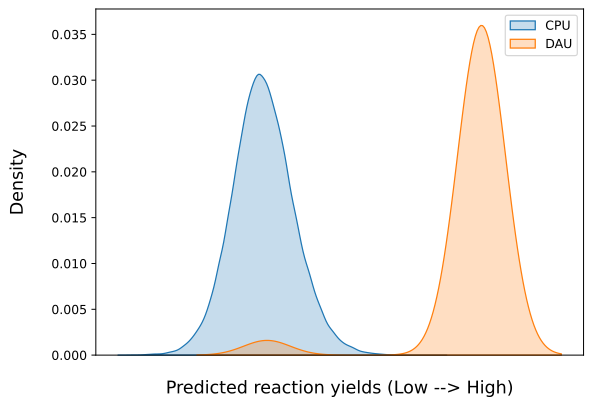}
  \vspace{20pt}
  \caption{Comparative analysis of the predicted reaction yield distributions for 100,000 randomly selected data points simulated by a CPU-based model (blue) versus the 24 unique results obtained from DAU-based model (orange).} 
 \label{fig:density_plot}
\end{figure}

The benchmarks of searching time are obtained from a laptop and a DAU. 
We employed trained QUBO models, focusing on the Negishi reaction with a reaction encoding of reac\_only. A random Negishi reaction was chosen for the inference process to identify the optimal reaction conditions for achieving higher yields. The dataset comprises 111 reagents and 24 solvents, resulting in a potential combination of 1,871,327,836, as calculated by $(\binom{111}{1} + \binom{111}{2} + \binom{111}{3} + \binom{111}{4}) * (\binom{24}{0} + \binom{24}{1} + \binom{24}{2})$ combinations for reaction conditions. Predicting all these values would be extremely time-consuming; hence, a hundred thousand data points were randomly sampled for inference using a laptop, which took about 20 minutes to complete. This suggests that predicting all possible combinations would take approximately 260 days. The optimal result obtained from the sampled points was then compared with the solutions suggested by the DAU, which ran for 10 seconds. The DAU model eventually returned 24 feasible reaction conditions, of which the reaction yields from 23 were higher than the optimal result from the sampled hundred thousand conditions. As shown in Figure \ref{fig:density_plot}, the outcomes produced by the DAU-based model significantly surpass those of a random selection from a set of 100,000 possible combinations, highlighting the DAU's exceptional capability in executing combinatorial optimization tasks. Moreover, the significant improvement in inference time when transitioning from a laptop to a DAU underscores the necessity of using a DAU for suggesting reaction conditions in systems with a large chemical space.

\section{Conclusions}
In this study, we constructed ML-based and DAU-based QUBO models to predict chemical reaction yield from reaction conditions. Results from subsets of high-throughput experimentation (HTE) datasets revealed that the performance of QUBO model is comparable to classical machine learning (ML) methods. In the extrapolation study, we trained our model on three out of four selected subsets and tested it on the remaining subset. Models can accurately predict top reaction yields in the s3 and s9 subsets with high correlation to true yields. However, QUBO models show inferior extrapolation in other subsets. To test model performance on more complex problems, we trained ML-based QUBO models using the Reaxys dataset and observed improvement with low encoding dimensions. These two pieces of evidence show the QUBO model can depict the hidden rules of chemical reactions. Furthermore, we investigated model performance in active learning, where limited data availability requires the model to determine optimal reaction conditions which can lead to higher yields from a small training set. Our results suggested that QUBO models captured the optimal reaction conditions corresponding to the highest yield rates in 10 iterations. However, improvements plateau after approximately 5 and 8 iterations in ML-based and DAU-based models, respectively. Building on the preliminary success of using DAU to optimize reaction conditions with significant speedup, this approach demonstrates considerable potential despite certain challenges. A critical factor is the accurate encoding of reaction conditions and chemical structures into QUBO form, which significantly impacts model performance. This work highlights the successful integration of ML and DAU for improved reaction yield prediction and condition optimization, emphasizing its promise for future research and applications.

\section*{Acknowledgements}
The authors thank Tzu-Lan Yeh and Yi-Shu Tu for the helpful discussions.

\bibliographystyle{unsrtnat}
\bibliography{bibfile}

\setcounter{figure}{0}
\setcounter{table}{0}
\renewcommand{\thefigure}{S\arabic{figure}}
\renewcommand{\thetable}{S\arabic{table}}
\renewcommand{\bibnumfmt}[1]{[S#1]}
\renewcommand{\citenumfont}[1]{S#1}

\newpage
\section*{Supporting Information}
\begin{figure}[H]
  \centering
  \includegraphics[width=0.99\linewidth]{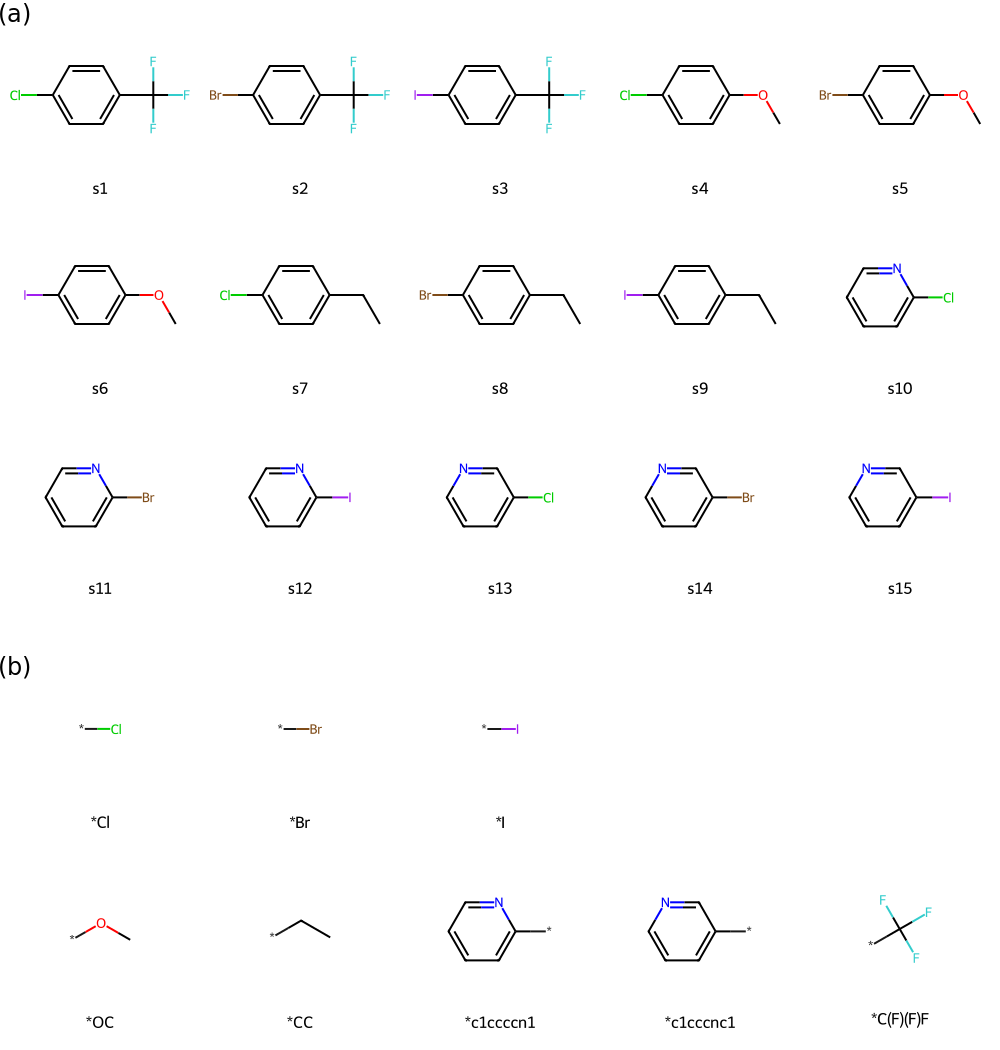}
  \vspace{20pt}
  \caption{(a) Chemical structures of substrates in the C-N cross-coupling dataset. (b) Substructures that can be used to construct the substrates, where each substrate is formed by combining a substructure from the top row with another from the bottom row.}
  \label{fig:cn-processed_substrate} 
\end{figure}

\newpage
\begin{figure}[H]
  \centering
  \includegraphics[width=\linewidth]{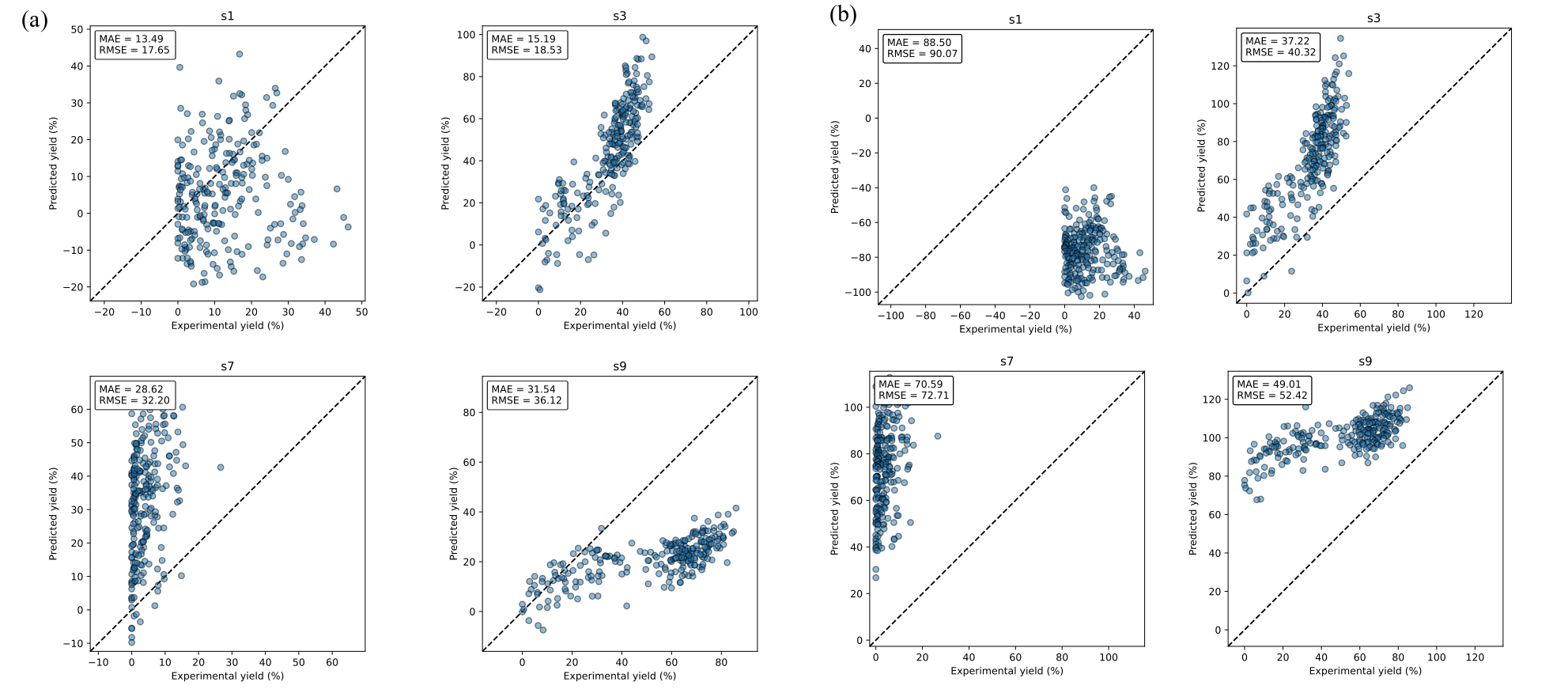}
  
  \caption{Results for the models trained on three subsets selected from s1, s3, s7, and s9, evaluated on the held-out test set of the C-N cross-coupling dataset via  (a) ML-based and (b) DAU-based model.}
  \label{fig:extrapolation_parity_plot} 
\end{figure}

\begin{figure}[H]
  \centering
  \includegraphics[width=0.8\linewidth]{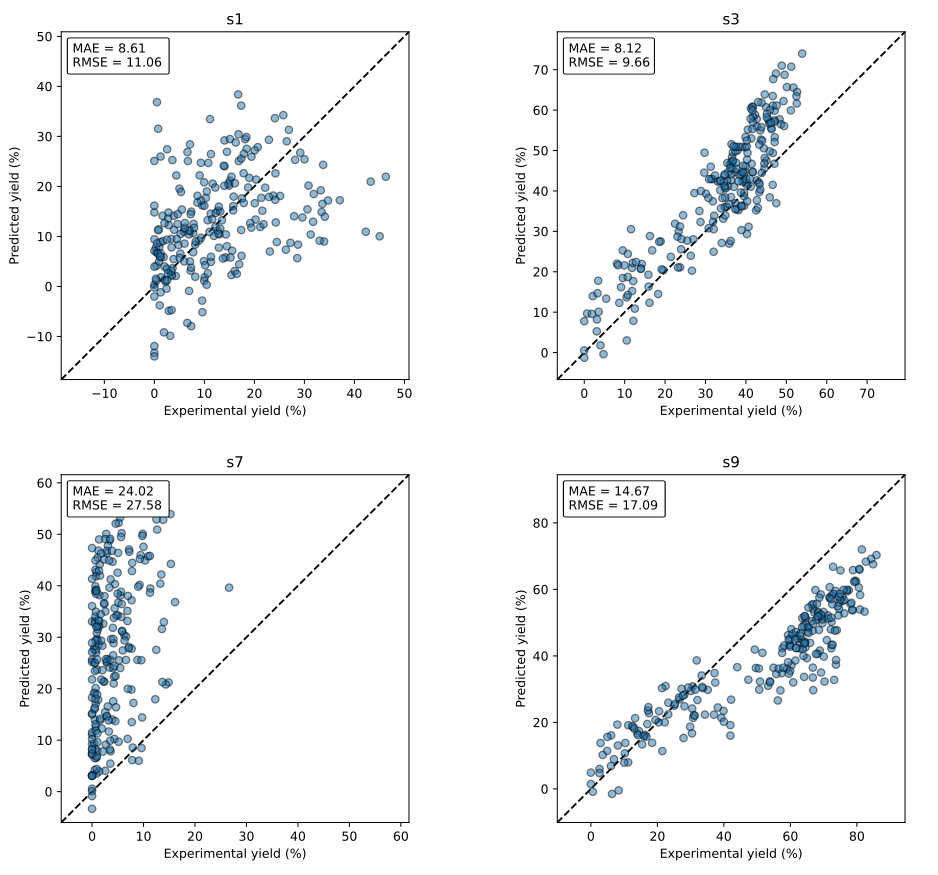}
  \caption{Results for the models trained on all subsets except for the held-out test set from the C-N cross-coupling dataset.}
  \label{fig:extrapolation_parity_plot_1} 
\end{figure}

\begin{figure}[H]
  \centering
  \includegraphics[width=0.8\linewidth]{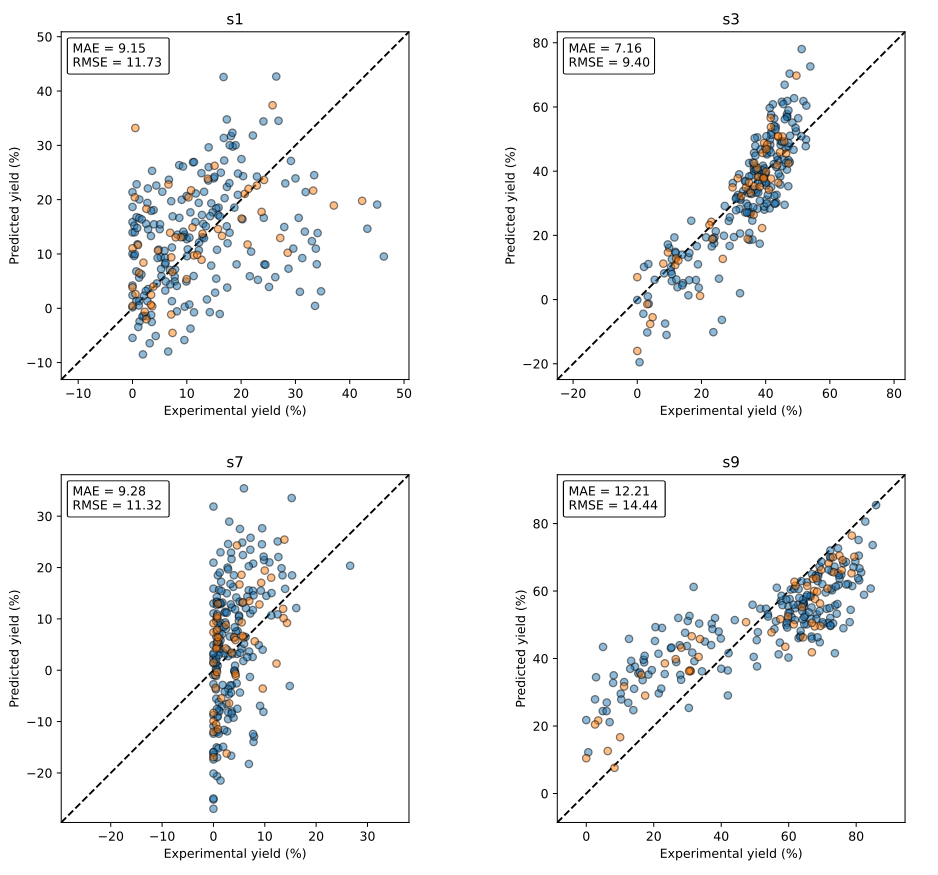}
  \caption{Results for the models trained on three subsets selected from s1, s3, s7, and s9, with an additional 50 data points from the held-out test set included in the training set, evaluated on the C-N cross-coupling dataset's held-out test set. Orange points represent the test data included in the training set; all other points belong to the test set.}
  \label{fig:extrapolation_parity_plot_2} 
\end{figure}

\begin{figure}[H]
  \centering
  \includegraphics[width=0.8\linewidth]{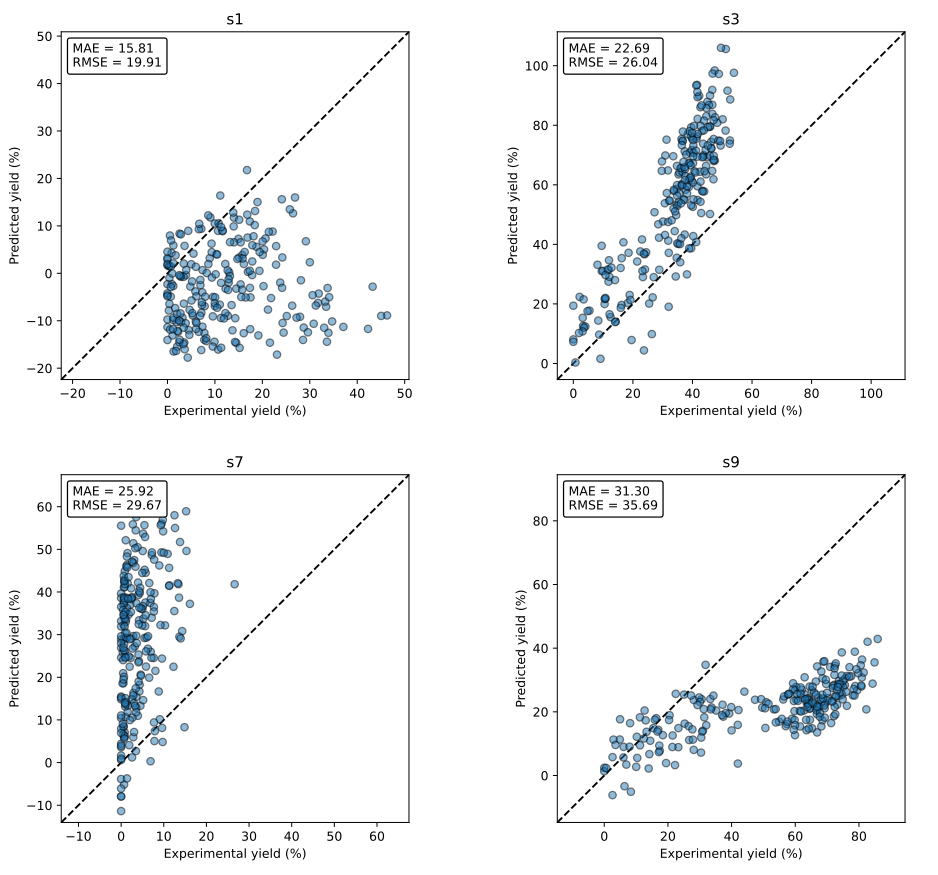}
  \caption{Results for the MLP models trained on three subsets selected from s1, s3, s7, and s9, evaluated on the held-out test set of the C-N cross-coupling dataset.}
  \label{fig:extrapolation_parity_plot_MLP} 
\end{figure}

\begin{figure}[H]
  \centering
  \includegraphics[width=0.8\linewidth]{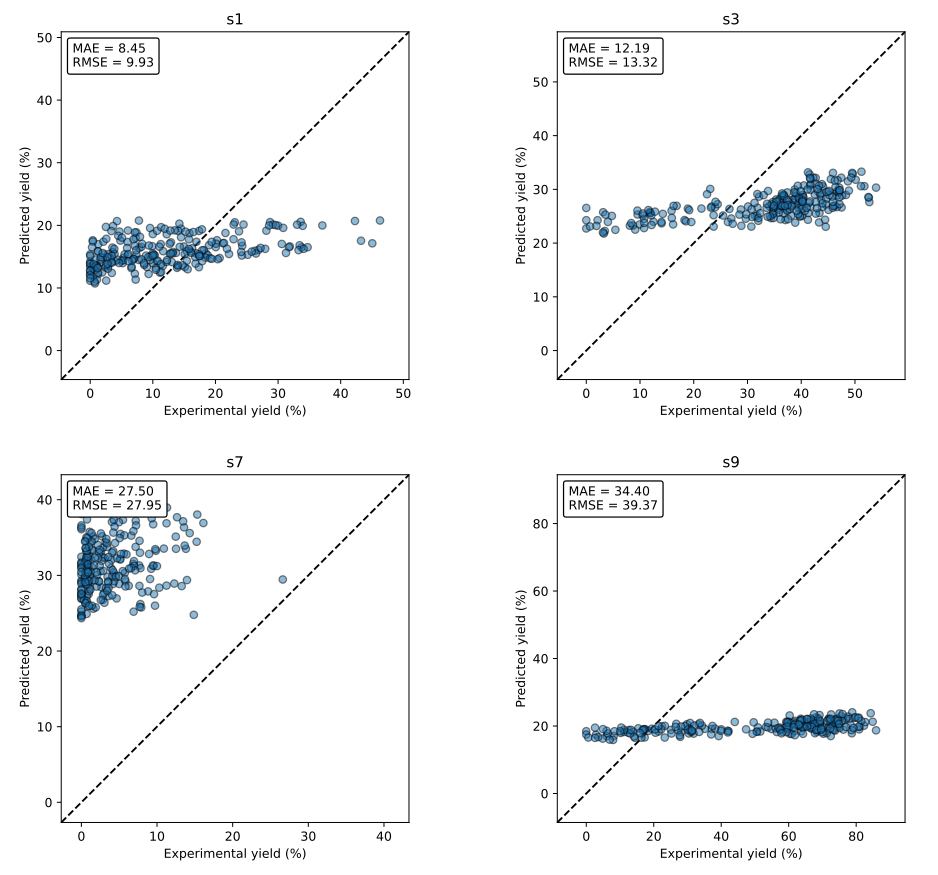}
  \caption{Results for the SVM models trained on three subsets selected from s1, s3, s7, and s9, evaluated on the held-out test set of the C-N cross-coupling dataset.}
  \label{fig:extrapolation_parity_plot_svm} 
\end{figure}

\begin{figure}[H]
  \centering
  \includegraphics[width=0.8\linewidth]{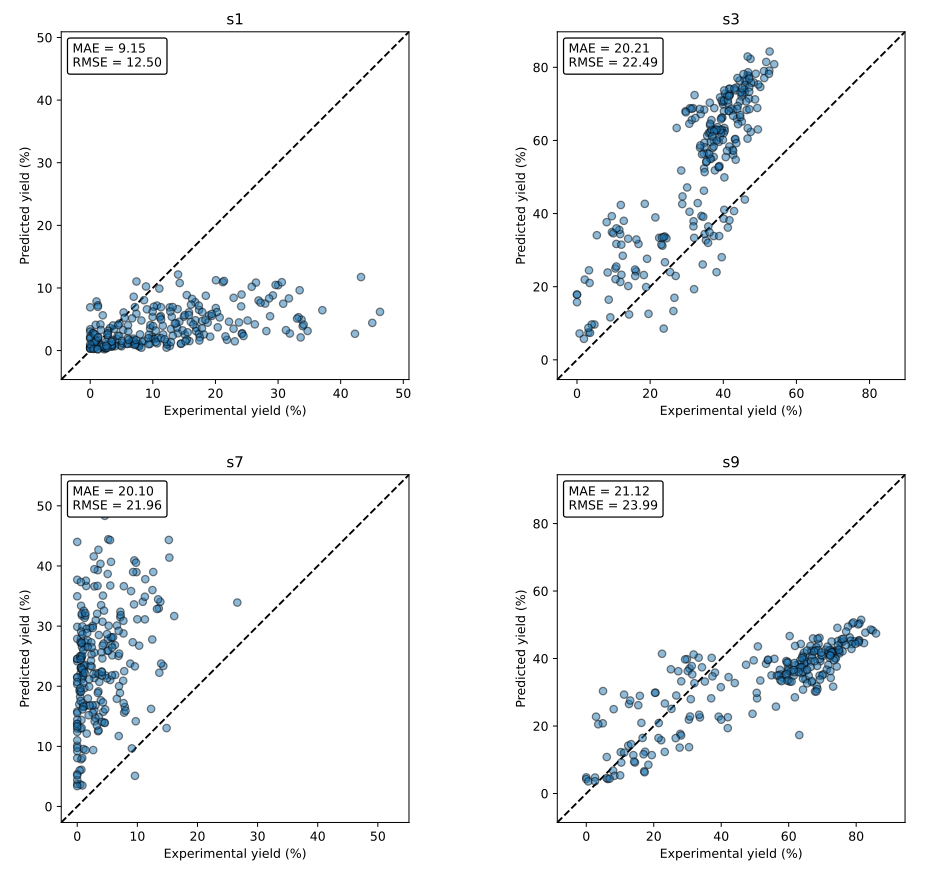}
  \caption{Results for the RF models trained on three subsets selected from s1, s3, s7, and s9, evaluated on the held-out test set of the C-N cross-coupling dataset.}
  \label{fig:extrapolation_parity_plot_random_forest} 
\end{figure}

\newpage
\begin{table}[H]
\caption{Testing MAE for models trained on Buchwald-Hartwig C-N cross-coupling reactions, using various combinations of reaction encoding approaches and fingerprints, from the Reaxys database. The temperatures are either not specified, provided in evenly distributed bins, or presented in a binary format.}
\begin{center}
\begingroup
\renewcommand{\arraystretch}{1.3} 
\resizebox{\textwidth}{!}{%
\begin{tabular}{lccccccccc} 
\toprule
& \multicolumn{3}{c}{Evenly distributed bins
} & \multicolumn{3}{c}{Binary encoding
} & \multicolumn{3}{c}{No contribution
} \\
\cmidrule(lr){2-4} \cmidrule(lr){5-7} \cmidrule(lr){8-10}
 & MACCS & Avalon & ECFP4 & MACCS & Avalon & ECFP4 & MACCS & Avalon & ECFP4 \\
\midrule
reac\_prod & 14.57 & 33.02 & 30.64 & 14.46 & 36.34 & 30.97 & 14.97 & 30.83 & 28.79 \\
reac\_only & 13.59 & 17.70 & 16.25 & 13.55 & 17.50 & 17.27 & 13.71 & 17.63 & 16.24 \\
prod\_only & 13.37 & 17.84 & 16.05 & 13.53 & 18.20 & 16.68 & 13.81 & 17.35 & 16.13 \\
reac\_diff & 13.59 & 31.69 & 21.83 & 13.67 & 35.61 & 23.87 & 13.75 & 37.84 & 22.92 \\
prod\_diff & 13.76 & 31.78 & 22.27 & 14.01 & 29.22 & 23.09 & 14.10 & 32.69 & 21.04 \\
diff\_only & 14.15 & 15.20 & 14.37 & 14.35 & 15.25 & 14.79 & 14.59 & 15.83 & 14.73 \\
\bottomrule
\end{tabular}%
}
\endgroup
\end{center}
\label{table:buchwald-hartwig}
\end{table}

\begin{table}[H]
\caption{Testing MAE for models trained on Suzuki reactions, using various combinations of reaction encoding approaches and fingerprints, from the Reaxys database. The temperatures are either not specified, provided in evenly distributed bins, or presented in a binary format.}
\begin{center}
\begingroup
\renewcommand{\arraystretch}{1.3} 
\resizebox{\textwidth}{!}{%
\begin{tabular}{lccccccccc} 
\toprule
& \multicolumn{3}{c}{Evenly distributed bins
} & \multicolumn{3}{c}{Binary encoding
} & \multicolumn{3}{c}{No contribution
} \\
\cmidrule(lr){2-4} \cmidrule(lr){5-7} \cmidrule(lr){8-10}
 & MACCS & Avalon & ECFP4 & MACCS & Avalon & ECFP4 & MACCS & Avalon & ECFP4 \\
\midrule
reac\_prod & 14.81 & 40.87 & 46.17 & 15.38 & 44.06 & 46.38 & 14.64 & 43.54 & 45.13 \\
reac\_only & 14.00 & 18.77 & 19.39 & 14.19 & 18.16 & 18.91 & 14.22 & 18.17 & 19.24 \\
prod\_only & 13.97 & 19.36 & 17.92 & 13.92 & 19.64 & 18.58 & 14.00 & 18.92 & 17.93 \\
reac\_diff & 14.51 & 41.97 & 31.75 & 14.40 & 46.72 & 33.23 & 14.67 & 40.62 & 30.80 \\
prod\_diff & 14.32 & 40.95 & 27.85 & 14.30 & 40.37 & 28.63 & 14.25 & 42.27 & 28.11 \\
diff\_only & 14.29 & 17.26 & 14.55 & 14.32 & 17.18 & 14.93 & 14.46 & 17.16 & 14.91 \\
\bottomrule
\end{tabular}%
}
\endgroup
\end{center}
\label{table:suzuki}
\end{table}

\begin{table}[H]
\caption{Testing MAE for models trained on Negishi reactions, using various combinations of reaction encoding approaches and fingerprints with different bit sizes (128, 256, 512, 1024, and 2048), from the Reaxys database. The temperatures are not used as input in these models.}
\begin{center}
\begingroup
\renewcommand{\arraystretch}{1.3} 
\resizebox{\textwidth}{!}{%
\begin{tabular}{lcccccccccc} 
\toprule
& \multicolumn{5}{c}{Avalon
} & \multicolumn{5}{c}{ECFP4
} \\
\cmidrule(lr){2-6} \cmidrule(lr){7-11}
 & 128 & 256 & 512 & 1024 & 2048 & 128 & 256 & 512 & 1024 & 2048 \\
\midrule
reac\_prod & 12.93  & 16.45 & 25.50 & 60.85 & 121.03 & 14.22 & 15.07 & 15.49 & 19.77 & 20.28 \\
reac\_only & 11.65  & 12.43 & 14.74 & 21.66 & 31.65  & 11.90 & 11.97 & 12.65 & 12.45 & 12.43 \\
prod\_only & 11.35  & 12.83 & 14.01 & 25.66 & 30.17  & 11.38 & 11.19 & 12.07 & 12.23 & 11.83 \\
reac\_diff & 13.73  & 16.84 & 27.34 & 46.58 & 90.73  & 13.08 & 13.62 & 14.99 & 15.52 & 17.20  \\
prod\_diff & 12.97  & 15.12 & 25.15 & 49.63 & 77.23  & 12.98 & 13.66 & 13.41 & 14.54 & 15.27 \\
diff\_only & 11.94  & 12.76 & 13.76 & 14.36 & 19.67  & 11.94 & 11.16 & 10.98 & 11.30 & 11.58 \\
\bottomrule
\end{tabular}%
}
\endgroup
\end{center}
\label{table:negishi_various_bits}
\end{table}

\newpage 
\end{document}